\documentclass{aastex63}
\usepackage{comment}
\usepackage{lineno}
\usepackage{array,multirow}
\usepackage{amsmath,accents,mathtools} % or simply amstext

\shorttitle{Rotational velocities of BSSs in NGC 3201}
\shortauthors{Billi et al.}
\graphicspath{{./}{figures/}}
\begin{document}

\title{Fast rotating Blue Straggler Stars in the Globular Cluster NGC
  3201 \footnote{Based on observations collected at the Clay Magellan
  Telescope, located at Las Campanas Observatory (Chile)}}

\correspondingauthor{Corresponding author}
\email{alex.billi2@unibo.it}

\author[0000-0002-3810-7343]{Alex Billi}
\affil{Dipartimento di Fisica e Astronomia, Universit\`a di Bologna, Via Gobetti 93/2 I-40129 Bologna, Italy}
\affil{INAF-Osservatorio di Astrofisica e Scienze dello Spazio di Bologna, Via Gobetti 93/3 I-40129 Bologna, Italy}

\author[0000-0002-2165-8528]{Francesco R. Ferraro}
\affil{Dipartimento di Fisica e Astronomia, Universit\`a di Bologna, Via Gobetti 93/2 I-40129 Bologna, Italy}
\affil{INAF-Osservatorio di Astrofisica e Scienze dello Spazio di Bologna, Via Gobetti 93/3 I-40129 Bologna, Italy}

\author[0000-0001-9158-8580]{Alessio Mucciarelli}
\affil{Dipartimento di Fisica e Astronomia, Universit\`a di Bologna, Via Gobetti 93/2 I-40129 Bologna, Italy}
\affil{INAF-Osservatorio di Astrofisica e Scienze dello Spazio di Bologna, Via Gobetti 93/3 I-40129 Bologna, Italy}

\author[0000-0001-5613-4938]{Barbara Lanzoni}
\affil{Dipartimento di Fisica e Astronomia, Universit\`a di Bologna, Via Gobetti 93/2 I-40129 Bologna, Italy}
\affil{INAF-Osservatorio di Astrofisica e Scienze dello Spazio di Bologna, Via Gobetti 93/3 I-40129 Bologna, Italy}

\author[0000-0002-5038-3914]{Mario Cadelano}
\affil{Dipartimento di Fisica e Astronomia, Universit\`a di Bologna, Via Gobetti 93/2 I-40129 Bologna, Italy}
\affil{INAF-Osservatorio di Astrofisica e Scienze dello Spazio di Bologna, Via Gobetti 93/3 I-40129 Bologna, Italy}

\author[0000-0002-3148-9836]{Lorenzo Monaco}
\affil{Instituto de Astrofísica, Facultad de Ciencias Exactas, Universidad Andres Bello, Sede
Concepcion, Talcahuano, Chile}

\author[0000-0002-3856-232X]{Mario Mateo}
\affil{Department of Astronomy, University of Michigan, 1085 S. University, Ann Arbor, MI 48109,USA}

\author[0000-0002-4272-263X]{John I. Bailey III}
\affil{Department of Physics, UCSB, Santa Barbara, CA 93016, USA}
 
\author[0000-0002-3887-6185]{Megan Reiter}
\affil{Department of Physics and Astronomy, Rice University, 6100 Main St., Houston, 77005-1827, Texas, USA}

\author[0000-0002-7157-500X]{Edward W. Olszewski}
\affil{Steward Observatory, The University of Arizona, 933 N. Cherry Avenue, Tucson, AZ 85721}

\begin{abstract}
We used high resolution spectra acquired at the Magellan Telescope to
measure radial and rotational velocities of approximately 200 stars in
the Galactic globular cluster NGC 3201. The surveyed sample includes
Blue Stragglers Stars (BSSs) and reference stars in different
evolutionary stages (main sequence turn-off, sub-giant, red giant and
asymptotic giant branches). The average radial velocity value
($\langle V_r\rangle = 494.5 \pm 0.5$ km s$^{-1}$) confirms a large
systemic velocity for this cluster and was used to distinguish 33
residual field interlopers. The final sample of member stars counts 67
BSSs and 114 reference stars. Similarly to what is found in other
clusters, the totality of the reference stars has negligible rotation
($<20$ km s$^{-1}$), while the BSS rotational velocity distribution
shows a long tail extending up to $\sim 200$ km s$^{-1}$, with 19 BSSs
(out of 67) spinning faster than 40 km s$^{-1}$. This sets the
percentage of fast rotating BSSs to $\sim 28\%$. Such a percentage is
roughly comparable to that measured in other loose systems ($\omega$
Centauri, M4 and M55) and significantly larger than that measured in
high-density clusters (as 47 Tucanae, NGC 6397, NGC 6752 and
M30). This evidence supports a scenario where recent BSS formation
(mainly from the evolution of binary systems) is occurring in
low-density environments. We also find that the BSS rotational
velocity tends to decrease for decreasing luminosity and surface
temperature, similarly to what is observed in main sequence stars. Hence,
further investigations are needed to understand the impact of BSS
internal structure on the observed rotational velocities.
\end{abstract}

\keywords{Blue Straggler Stars --- Globular clusters: individual (NGC 3201) --- Spectroscopy}

\section{Introduction} 
\label{sec:intro}
Blue Straggler Stars (BSSs) are an ``exotic'' stellar population
observed in different stellar environments, such as globular clusters
(GCs; \citealp{sandage+53, ferraro+97, ferraro+99, ferraro+03,
  ferraro+09, piotto+04, lanzoni+07, leigh+07, dalessandro+08,
  moretti+08}), open clusters \citep{mathieu+09, geller+11,
  gosnell+14}, dwarf spheroidal galaxies \citep{momany+07,
  mapelli+09}, the Milky Way halo and bulge \citep{preston+00,
  clarkson+11}.  In the color-magnitude diagram (CMD), BSSs lie on a
bluer extension of the main sequence (MS) and have luminosities
brighter than the MS turn-off. Hence, BSSs are more massive than MS
stars (M $\sim$ 1.2-1.6 M$_\odot$; \citealp{shara+97, gilliland+98,
  demarco+05, fiorentino+14, raso+19}). Because of their relatively
large mass, BSSs are crucial ``gravitational test particles'' to probe
internal dynamics of GCs \citep[e.g.,][]{ferraro+09, ferraro+12,
  ferraro+20}. In this respect \citet{ferraro+12} introduced the
concept of ``dynamical clock'', an empirical method to measure the
level of dynamical evolution of star clusters by using the BSS radial
distribution. This concept was further refined by the definition of
the $A^+_{rh}$ parameter \citep{alessandrini+16, lanzoni+16}, which
quantifies the level of BSS central segregation with respect to a
lighter (reference) population, and has been successfully adopted to
rank a large sample of Galactic \citep{ferraro+18a, ferraro+23a} and
extra-Galactic star clusters \citep{ferraro+19, dresbach+22} in terms
of their dynamical stage.

The formation and evolutionary processes of BSSs are still not
completely understood, although several progresses in the
observational characterization and theoretical modeling of these
stars have been made in the recent years, especially in open clusters
(see, e.g. \citealt{mathieu+09, perets+09, chatte+13, leiner+21,
  jiang22, reinoso+22, cordoni+23}, and references therein).  Three
main formation scenarios have been suggested so far: {\it (i)} mass
transfer activity in binary systems \citep{mccrea64}, where a
companion star transfers mass and angular momentum to the accreting
proto-BSS, {\it (ii)} direct collisions between two or more stars
\citep{hills+76}, and {\it (iii)} mergers/collisions induced by
dynamical/stellar evolution in triple systems (see
\citealp{andronov+06,perets+09}). The latter is thought to play a
major role in the field and in open clusters, and it naturally
accounts for the observed period-eccentricity distribution of BSS
binaries (\citealp{perets+09}). The mass-transfer scenario is expected
to be dominant in low density systems (see, e.g., \citealp{mateo+90,
  sollima+08, mathieu+09}), and photometric confirmation of this
formation channel has been recently obtained in open clusters through
the detection of ultraviolet emission from hot white dwarf companions
(the peeled donor stars) to a few BSSs \citep{gosnell+14, gosnell+15}.
A potential spectroscopic signature of BSS formation through
mass-transfer activity has been also identified in a few BSSs in
Galactic GCs. In fact, at odds with collisional BSSs
\citep{lombardi+95}, mass-transfer BSSs are predicted to show carbon
and oxygen depletion on their surface \citep{sarna+96}, due to the
accretion of material coming from the innermost layers of the donor
star, which are expected to be partially processed by the CNO burning
cycle.  \citet{ferraro+06a} and \citet{lovisi+13a} detected carbon and
oxygen depletion in a sub-sample of BSSs in 47 Tucanae and M30,
respectively.  On the other side, sub-populations of collisional BSSs
have been possibly identified in the CMD of a few post core collapsed
GCs \citep{ferraro+09, portegies19, dalessandro+13, beccari+19,
  cadelano+22}, where they appear all aligned along a narrow blue
sequence, well separated from the remaining population of redder
BSSs. The formation of these collisional BSSs is thought to be
promoted by the enhanced collisional activity occurring in the cluster
core during the core collapse and the post core collapse stages, when
the central density substantially increases.

The study of rotational velocities can add an important piece to the
BSS formation puzzle, providing additional constraints to their origin
and evolutionary processes. From a theoretical point of view, large
rotational velocities are expected at birth for both mass-transfer
BSSs \citep{sarna+96} and collisional BSSs \citep{benz+87}, but some
braking mechanisms (such as magnetic braking and disk locking) should
then intervene to slow down these stars with efficiencies and time
scales that are not fully understood yet \citep[e.g.,][]{leonard+95,
  sills+05}.  Recent results in open clusters suggest that magnetic
braking is efficient enough to slow down mass-transfer BSSs in
timescales shorter than 1 Gyr \citep{leiner+18}. In fact, the relation
between the measured BSS rotation periods and ages (derived from the
cooling time of the white dwarf companion) is well reproduced by the
same models used to describe the spin-down rate of single, low-mass MS
stars (which, indeed, share the same structure in terms of convective
envelopes and magnetic fields; \citealt{gallet+15}).
%thus providing a braking time of the order of 1 Gyr.
This result indicates that the BSS rotation rate can potentially be
used as gyro-chronometer to measure the time since the end of
mass-transfer.  Additional complexity is introduced by phenomena like
the synchronization of close binaries (e.g., Hut 1981), which tends to
spin up the stars but is very hard to constrain observationally, since
it requires dedicated multi-epoch observations in both photometry and
spectroscopy (see, e.g., \citealp{meibom+06, lurie+17, leiner+19}, and
references therein).
 
In this context, several years ago, our group started an extensive
high-resolution spectroscopic survey at large ground-based telescopes
for eight Galactic GCs with different structural parameters, with the
aim to provide the rotational velocities of a representative sample of
BSSs formed and evolved in different environments.  The results
obtained for six clusters have been already published: see
\citet{ferraro+06a} for 47 Tucanae, \citet{lovisi+10, lovisi+12,
  lovisi+13a, lovisi+13b} for M4, NGC 6397, M30, and NGC 6752,
respectively, and \citet{mucciarelli+14} for $\omega$ Centauri.  These
data, complemented with the results obtained for NGC 3201 (which are
presented in this paper), and for M55 (which will be discussed in
Billi et al. 2023, in preparation) have been recently used by
\citet{ferraro+23b} to demonstrate that the fraction of fast rotating
BSSs (with $v\sin(i)\ge 40$ km s$^{-1}$, with $i$ being the
inclination angle on the plane of the sky) anti-correlates with the
central density of the host system, decreasing from $\sim 40\%$ to
$\sim 5\%$ when the central density increases from $\log \rho_0=2.5$
to $\log \rho_0= 5$ (in units of $L_\odot/$pc$^3$), thus indicating
the tendency of fast spinning BSSs to populate low-density
environments.

The present paper is devoted to discuss the results obtained for 67
BSSs observed in NGC 3201 at a spectral resolution $R=18,000$, and
compare them to the rotational velocities measured by
\citet[][hereafter, SP14]{sp14} for 37 such objects in the same
cluster,
%using the Inamori Magellan Area Camera \& Spectrograph at Baade
%Magellan Telescope,
at $R=10,000$.  The paper is organized as follows: the observations
are presented in Section \ref{sec:obs}, the determinations of radial
velocities, atmospheric parameters, and rotational velocities are
discussed in Sections \ref{sec:rv_membership},
\ref{sec:atmospheric_parameters}, and \ref{sec:rotational_velocities},
respectively. Section \ref{sec:comparison_sp14} discusses the
comparison between our findings and those of SP14.  A summary of the
results and the conclusions of the work are provided in Section
\ref{sec:discussion}.
 
\begin{figure}[ht!]
\includegraphics[width=10.5cm, height=10.5cm]{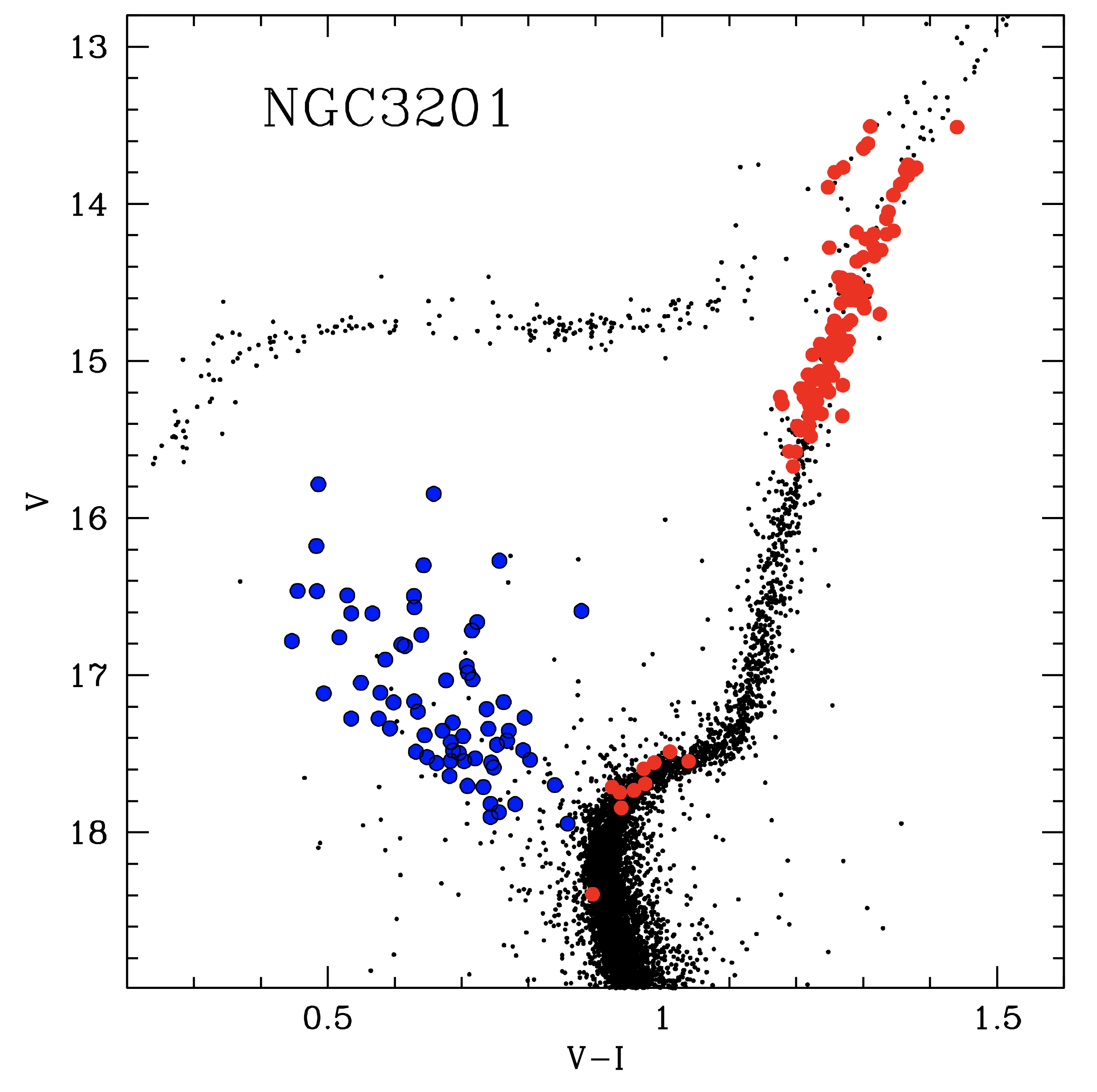}
\centering
\caption{Proper motion selected and differential reddening corrected
  CMD of NGC 3201 (black dots), with the spectroscopic targets
  discussed in the paper highlighted as filled colored circles: the 67
  BSSs are marked in blue, while the reference sample of MS, SGB, RGB
  and AGB stars are plotted in red. }
\label{fig:cmd_target}
\end{figure}

\section{Observations}
\label{sec:obs}
This work is based on stellar spectra acquired with the multi-object
spectrograph Michigan/Magellan Fiber System (M2FS, \citealp{mateo+12})
mounted on the Clay Magellan Telescope, located at Las Campanas
Observatory (Chile). M2FS is composed of two spectrographs fed by 128
fibers each. The observations have been performed during two nights
(on 2019, February 28 and March 2), adopting the Mgb$_{\rm Revb}$
configuration, which covers the spectral region between 5127 \AA\ and
5184 \AA, sampling the first two lines of the Mg triplet (at 5167.3
\AA\ and 5172.6 \AA) with a spectral resolution $R\sim 18,000$.  Two
different fiber configurations have been adopted to secure spectra for
approximately 200 targets in the cluster direction. A total of 6
exposures of 30 minutes each, and 10 exposures of 20 minutes have been
secured in the first and second night, respectively.

The BSS selection has been performed on the basis of high-resolution
ultraviolet catalogs obtained from HST observations for the central
regions of NGC 3201, and optical ground-based observations for the
external regions (see \citealt{ferraro+23a}).  For a proper comparison
with SP14, particular care has been devoted to include in the sample
most of the BSSs observed in that work (29 out 37). In addition, to
compare the BSS rotational velocities with those of ``normal'' cluster
stars, we simultaneously observed a significant sample of reference
objects distributed in different evolutionary stages. Due to the high
value of the seeing (larger than $1.50\arcsec$), 2 and 3 exposures
from the first and the second night, respectively, have been excluded
from the analysis.  As a result, we secured spectra of 67 BSSs, 9
sub-giant branch (SGB) stars, one MS star, 98 red giant branch (RGB)
stars, and 6 asymptotic giant branch (AGB) stars.  Figure
\ref{fig:cmd_target} shows the position of the observed sample in the
proper motion selected and differential reddening corrected CMD of NGC
3201, obtained from the photometric catalog of the Stetson database
\citep{stetson+19} by applying the method described in \citet[][see
  also \citealt{deras+23}]{cadelano+20}.

\section{Radial velocities and cluster membership}
\label{sec:rv_membership}
The raw spectra have been pre-reduced by following the standard
procedure including bias subtraction, flat field correction,
wavelength calibration and, finally, the 1D spectra extraction. The
median of the sky spectra secured during the observations has been
assumed as master-sky spectrum and used to remove the sky contribution
from each individual exposure. Then, for each target the
sky-subtracted spectra have been combined, providing the final
spectrum for the analysis.  The radial velocities of the sampled stars
have been measured by using the IRAF task \textit{fxcor}, which
performs a cross-correlation between the observed spectrum and a
template of known radial velocity \citep{tonry+79}. As templates we
adopted synthetic spectra computed with the code SYNTHE
\citep{sbordone+04, kurucz05} by adopting the last version of the
Kurucz/Castelli linelist for atomic and molecular transitions.  To
take into account the different stellar parameters and line strengths,
we defined a template for giant (RGB and AGB) stars, and another one
for BSSs and SGB/TO stars.  The model atmospheres have been computed
with the ATLAS9 code \citep{kurucz93, sbordone+04} under the
assumptions of local thermodynamic equilibrium (LTE) and
plane-parallel geometry, and adopting the new opacity distribution
functions by \citet{castelli+03} with no inclusion of
overshooting \citep{castelli+97}.

\begin{figure}[ht!]
\includegraphics[width=9cm, height=9cm]{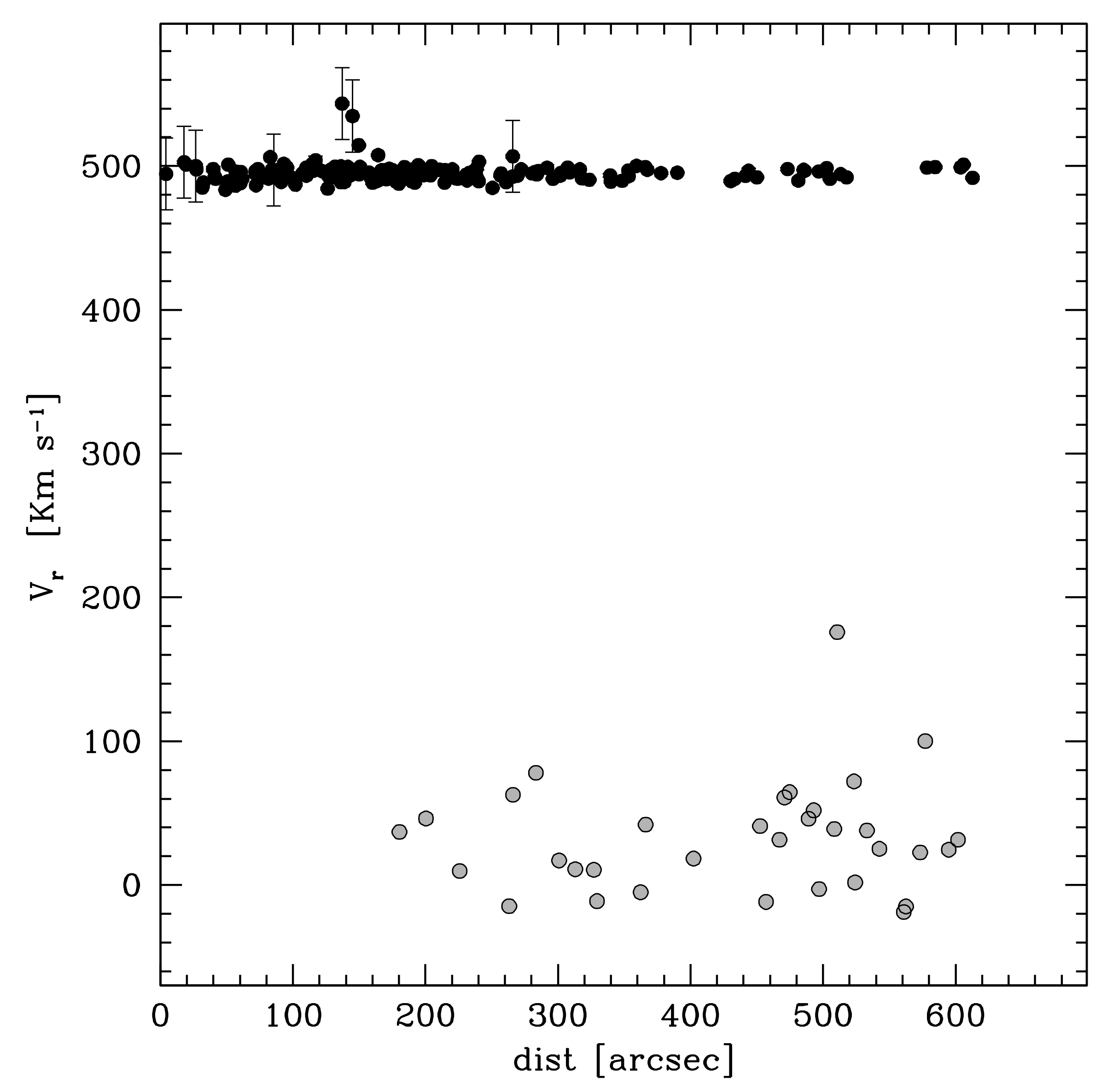}
\centering
\caption{Radial velocities (in km s$^{-1}$) of the target stars as a
  function of their distance from the cluster center (in
  arcseconds). The stars belonging to NGC 3201 are shown as black
  circles, while the objects belonging to the Galactic field are
  colored in grey.}
\label{fig:RV}
\end{figure}

Figure \ref{fig:RV} shows the distribution of the measured radial
velocities as a function of the distance from the cluster center. Here
we adopt the center of gravity obtained by B. Lanzoni et al. (2023, in
preparation): $\alpha_{J2000}$ = $10^h17^m36.82^s$ $\delta_{J2000}$ =
$-46^{\circ} 24\arcmin 44.9\arcsec$.  The acquired dataset extends
from $r=4\arcsec$, out to $\sim 630\arcsec$. The cluster population
(black circles) is clearly distinguishable as a narrow component
strongly peaked at the large systemic velocity of the cluster: $V_{\rm
  sys}= 494.5 \pm 0.5$ km s$^{-1}$ \citep{ferraro+18mikis}. This
allows a straightforward identification of the few field stars
included in the observed sample, which show much smaller radial
velocities (grey circles in Figure \ref{fig:RV}).  The few cluster
members showing radial velocities larger than the average are rapidly
rotating BSSs for which the measures are more uncertain because of the
significant deformation of the absorption lines (see Section
\ref{sec:rotational_velocities}).

\begin{figure}[ht!]
\includegraphics[width=9cm, height=9cm]{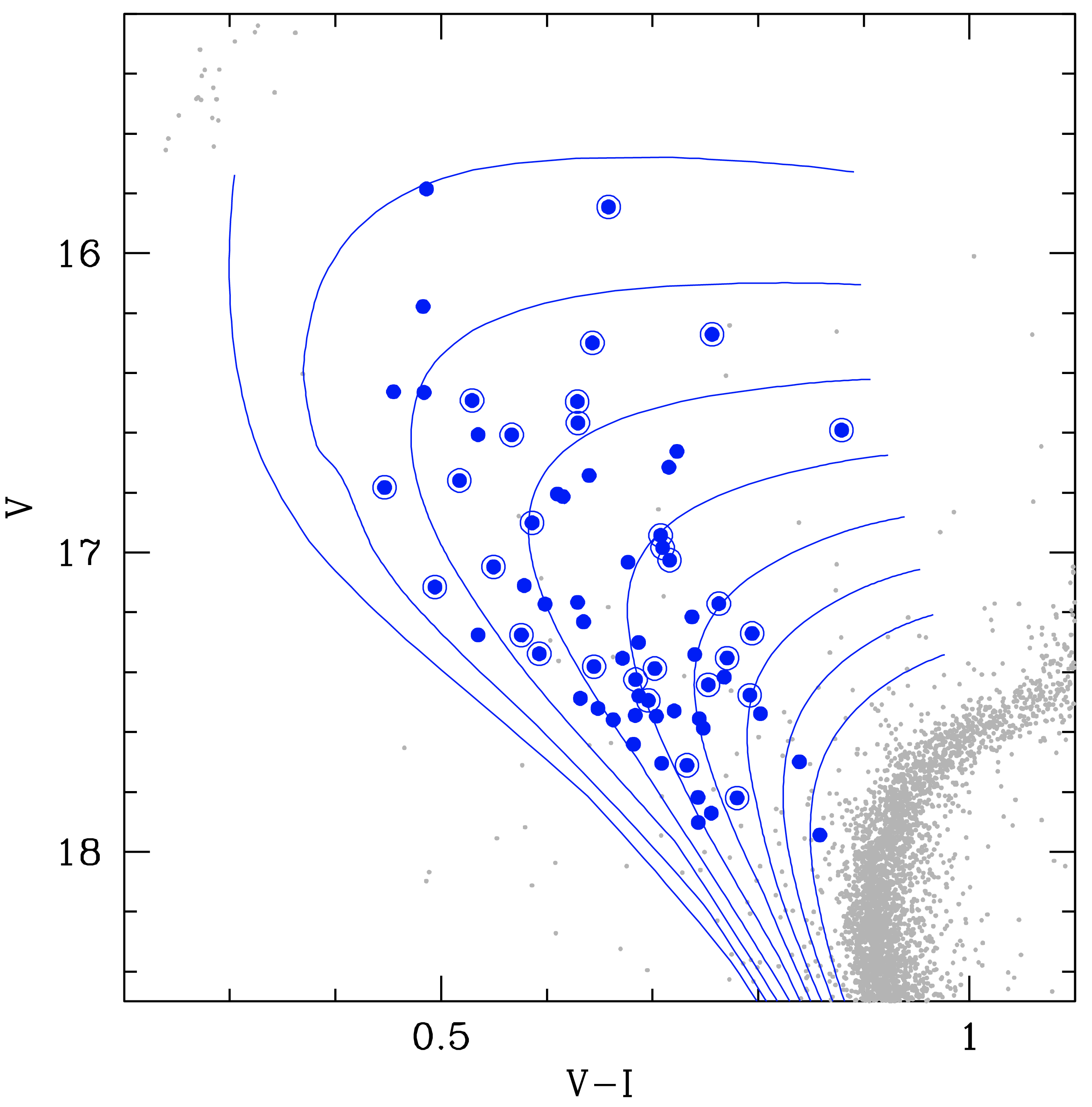}
\centering
\caption{CMD of NGC 3201 (grey dots) with the 67 surveyed BSSs marked
  as blue circles, and a set of BaSTI isochrones
  \citep{pietrinferni+21} with ages ranging from 1 to 9 Gyr
  overplotted as blue lines.  The BSSs in common with SP14 are
  encircled.}
\label{fig:isoch}
\end{figure}

%%%%%%%%%%%%%%%%%%%%%%%%%%%%%%%%%%%%%%%%%%%%%%%%----RISULTATI

\section{Atmospheric parameters}
\label{sec:atmospheric_parameters}
To estimate the effective temperature ($T_{\rm eff}$) and the surface
gravity ($\log g$) of the observed BSSs we have compared their CMD
location with a set of $\alpha-$enhanced BaSTI isochrones
\citep{pietrinferni+21}, computed by assuming the cluster metallicity
([Fe/H]$=-1.55$), and ages ranging between 1 and 9 Gyr.  The models
have been reported to the observed CMD (see blue lines in Figure
\ref{fig:isoch}) by adopting an apparent distance modulus $(m-M)_V =
14.20$ and a color excess $E(B-V) = 0.27$, in agreement with
\citet{harris+96}.  As apparent from the figure, the grid of
isochrones properly samples the portion of the CMD where BSSs are
distributed.  Hence, the needed values of $T_{\rm eff}$ and $\log g$
have been determined by projecting each BSS onto the closest
isochrone. The resulting effective temperatures vary between 6700 and
8700 K, while $\log g$ ranges between 3.6 and 4.3 dex.  An analogous
procedure, but using the 12 Gyr old isochrone, has been performed for
the reference stars.  It is worth emphasizing that these photometric
estimates of temperature and gravity are accurate enough to guarantee
a solid determination of the rotational velocity. In fact, the impact
of these two parameters on the measured rotation is modest. For
instance, even large variations in temperature and gravity ($\pm
300-500$ K in $T_{\rm eff}$ and $\pm$ 0.8 dex in $\log g$) produce a
small effect ($\sim 1$ km s$^{-1}$) in the measured value of $v
\sin(i)$.  In addition, we assumed a microturbulence velocity of 1 km
s$^{-1}$ for the BSSs, and 1.5 km s$^{-1}$ for the giant stars,
although different assumptions for this parameter have no impact on
the rotational velocity values.
 
\section{Rotational velocities}
\label{sec:rotational_velocities}
\begin{figure}[ht!]
\includegraphics[width=12cm, height=12cm]{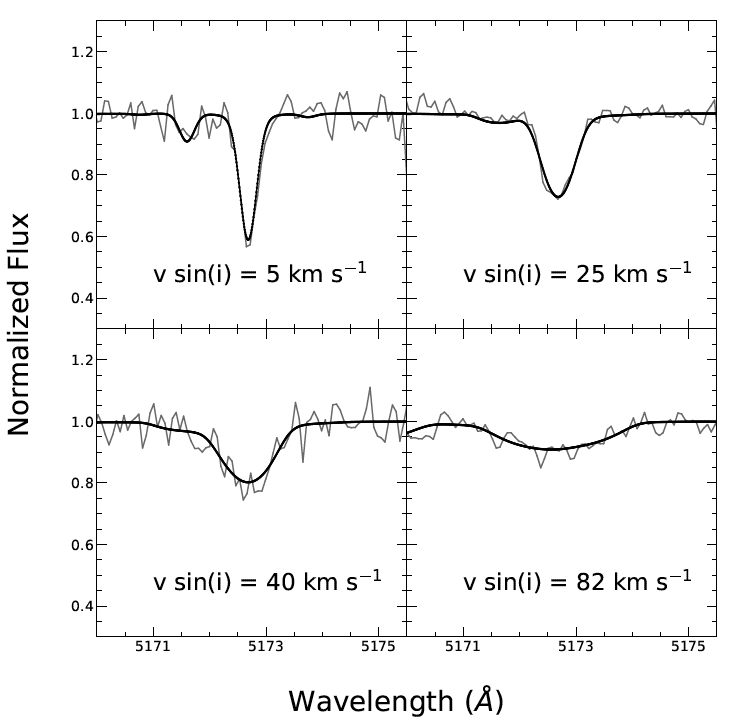}
\centering
\caption{Comparison between the observed spectra (grey lines) and the
  best-fit synthetic spectra (black lines) for 4 BSSs with different
  rotational velocities, ranging from 5 to 82 km s$^{-1}$.  }
\label{fig:vrot_oss_sint}
\end{figure}

The rotational velocities (projected on the plane of the sky) have
been calculated by using a $\chi^2$ minimization procedure between the
observed spectra, and a grid of synthetic spectra calculated for the
atmospheric parameters estimated as described in Section
\ref{sec:atmospheric_parameters}, and for different values of $v
\sin(i)$.  The rotational velocity has been measured from the two
first lines of the Mg triplet at 5167.3 and 5172.6 \AA. These two
features are strong enough to allow an accurate measure of $v \sin(i)$
also in the case of very large rotations (of the order of 100 km
s$^{-1}$).  For the sake of illustration, Figure
\ref{fig:vrot_oss_sint} shows the comparison between the observed and
the synthetic spectra of four stars with different rotational
velocities for the line at 5172.6 \AA. As can be seen, the absorption
line is still clearly visible also in the star rotating at more than
80 km s$^{-1}$.

The uncertainties in the measured rotational velocities have been
determined through Monte Carlo simulations.  For each star we
calculated a synthetic spectrum adopting the atmospheric parameters
and rotational velocity previously obtained.  Then, through the random
additions of Poissonian noise, we simulated 300 spectra with the same
signal to noise ratio of the observed one.  These have been analyzed
as described above, each time obtaining the best-fit rotational velocity
value.  The standard deviation of the derived velocity distribution
has been assumed as 1$\sigma$ uncertainty in the rotational velocity
of each star. The estimated uncertainties range from a few (2-5) km
s$^{-1}$ for slowly rotating stars, up to 15-20 km s$^{-1}$ for the
fast spinning targets (with rotation of the order of, or larger than,
100 km s$^{-1}$).

\begin{figure}[ht!]
\includegraphics[width=14cm, height=14cm]{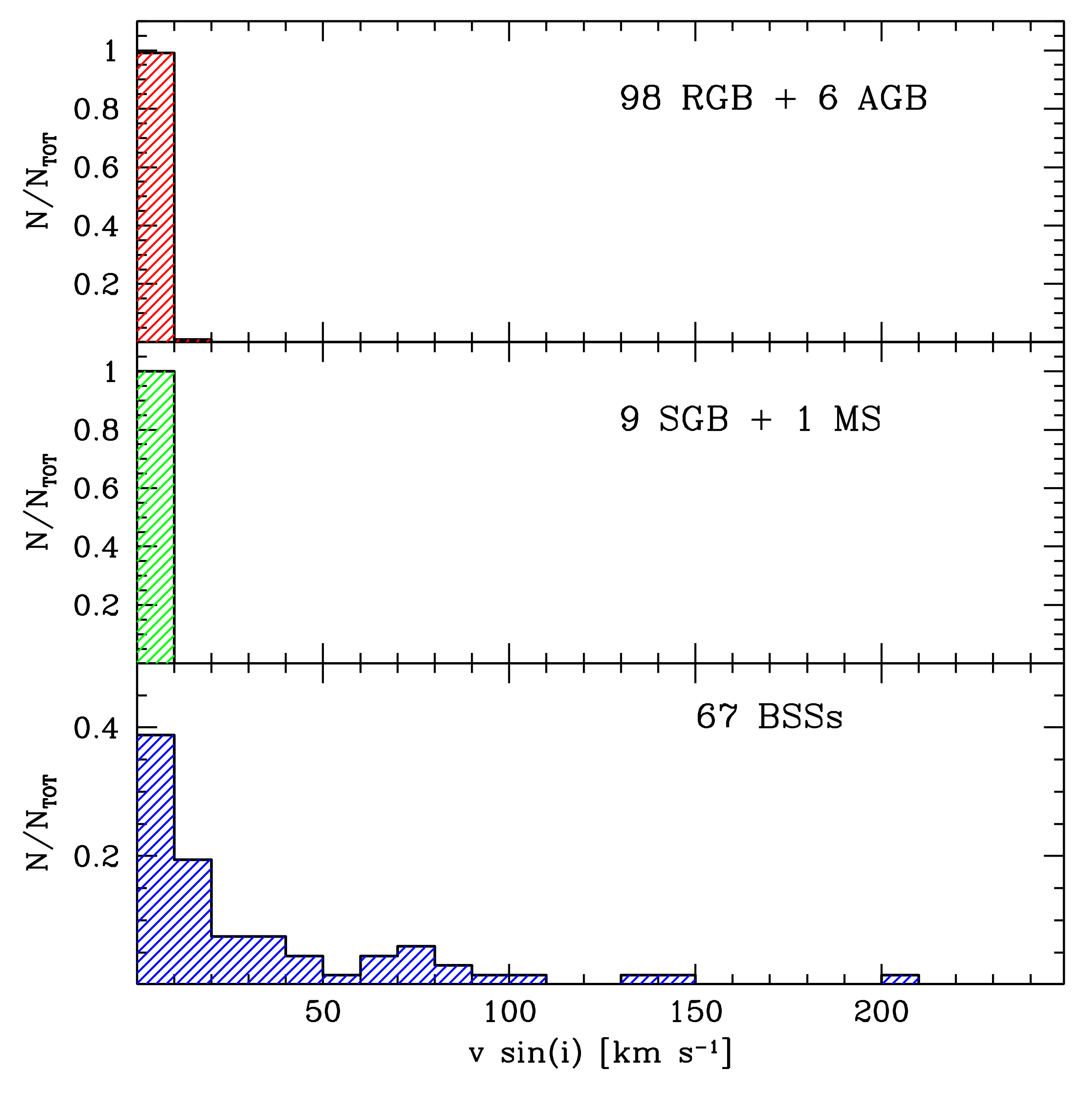}
\centering
\caption{Distributions of the rotational velocities measured for
  highly-evolved, giant stars (red histogram in the upper panel),
  mildly-evolved (SGB and MS) stars (green histogram in the central
  panel), and BSSs (blue histogram in the bottom panel).}
\label{fig:histo}
\end{figure}

As shown in Figure \ref{fig:histo}, the distributions of rotational
velocities obtained for the observed samples of BSSs and normal
cluster stars are strikingly different.  Almost all the highly evolved
stars (RGB and AGB targets), all the SGB stars and the MS star in the
sample show negligible rotation ($v \sin(i)<10$ km s$^{-1}$), in
agreement with the rotational velocities quoted in the literature for
cluster and field stars in similar evolutionary stages (see, e.g.,
\citealp{lucatello+03, cortes+09, demedeiros+14}). Indeed only one RGB
star has a rotation of 12 km s$^{-1}$, which could be due to an
interaction with a companion object (unfortunately our dataset is not
suitable to search for radial velocity variability, but this object
%(that could be considered a signature of binarity) and it
is certainly worth of further investigations).  Conversely, the
distribution of BSS rotational velocities appears to be much more
articulated, with a peak at low values followed by a long tail
extending up to very large rotations (up to 207 km
s$^{-1}$). Moreover, only less than $40\%$ of the entire sample shows
rotational velocities below 10 km s$^{-1}$, while 19 BSSs our 67
($\sim 28\%$ of the total) spin faster than 40 km s$^{-1}$ and,
according to the definition of \citet{ferraro+23b}, are classified as
fast rotators (FRs). Hence, this comparison clearly demonstrates that
BSSs display a very peculiar rotational velocity distribution with
respect to normal clusters stars.

%\begin{deluxetable*}{rrrrccccc}[ht!]
%\tablecaption{TABLE CAPTION.}
%\tablewidth{0pt}
%\renewcommand{\arraystretch}{1.2}
%\tablehead{
%\colhead{col1 } & \colhead{ col2 } & \colhead{ col3 } & \colhead{ col4}  &  \colhead{col5}  &
%\colhead{ col6} & \colhead{ col7}
%}
%\startdata
%    &   &  &   &  &  &   \\ 
%    &   &  &   &  &  &   \\ 
%%   &   &  &   &  &  &   \\
%\enddata
%\tablecomments{TABLE COMMENTS.}
%\label{tab:rv_atmparams_vrot}
%\end{deluxetable*}

\section{Comparison with SP14}
\label{sec:comparison_sp14}
As mentioned in the Introduction, SP14 determined the rotational
velocities of 37 BSSs in NGC 3201 (plus other BSSs in $\omega$
Centauri and NGC 6218) from the analysis of spectra acquired with the
Inamori Magellan Area Camera \& Spectrograph at the Baade Magellan
Telescope, at a spectral resolution $R=10,000$.  The rotational
velocity distribution derived by these authors is different from that
plotted in Figure \ref{fig:histo}: it shows a peak at 25-30 km
s$^{-1}$, which is not visible in our distribution, and only 6 stars
($\sim$ 19 \% of the total sample) spin faster than 40 km s$^{-1}$.
Since we have 29 BSSs in common with SP14, we can perform a proper
star-to-star comparison to understand the origin of the discrepancy.

\begin{figure}[ht!]
\includegraphics[width=16cm, height=7cm]{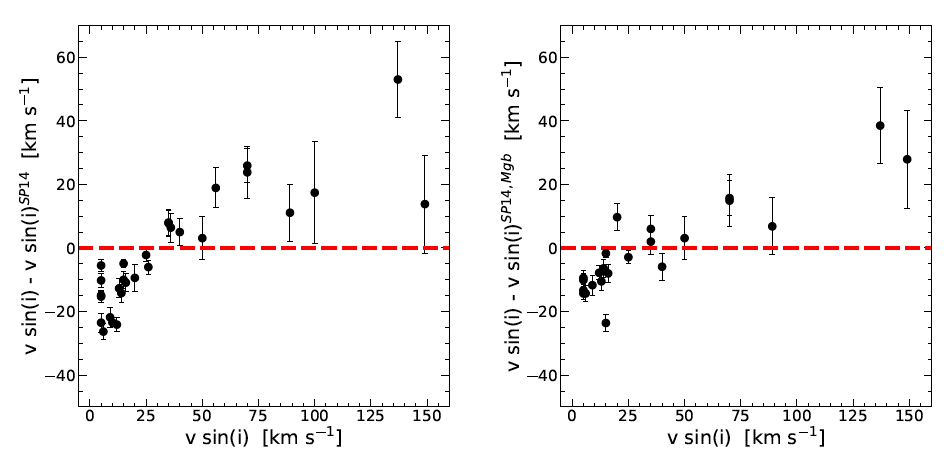}
\centering
\caption{\emph{Left panel:} Difference between the rotational
  velocities obtained in this work ($v \sin(i)$) and those determined
  by SP14 ($v \sin(i)^{\rm SP14}$) for the 29 BSSs in common, as a
  function of our measures. \emph{Right panel:} as in the left panel,
  but for the rotational velocities of SP14 computed from the Mgb
  triplet line alone ($v \sin(i)^{\rm SP14, Mgb}$).}
\label{fig:comparison_SP14}
\end{figure}

Figure \ref{fig:comparison_SP14} shows the difference between the
rotational velocities determined in this work and in SP14, as a
function of our measures, for the 29 BSSs in common.  As apparent, all
the stars for which we find $v \sin(i)< 30$ km s$^{-1}$ have larger
rotation in SP14 (in some extreme cases they are in excess by even
$\sim 25$ km s$^{-1}$), while the opposite is true for most of the
FRs.  This trend is very similar to that found by
\citet{mucciarelli+14} in the case of $\omega$ Centauri (see their
Figure 6), from the comparison between the measurements of SP14 and
those obtained from higher resolution ($R>18,000$) spectra acquired
with FLAMES at the ESO-VLT for 14 BSSs in common. Also taking into
account that the spectral resolution used in the present work and in
\citet{mucciarelli+14} is approximately the same and about twice that
of SP14, we conclude that the origin of the discrepancies observed in
NGC 3201 (Fig. \ref{fig:comparison_SP14}) and in $\omega$ Centauri
(Fig. 6 in \citealp{mucciarelli+14}) is the same.  The disagreement
detected at low rotational velocities can be ascribed to the lower
spectral resolution used in SP14, which naturally makes much more
difficult the measure of small rotation values because the line
profile is dominated by the instrumental broadening, instead of the
rotational broadening.

The difference in the high velocity regime is mainly ascribable to the
different absorption lines used in the SP14 analysis.  In fact, in
addition to the Mgb triplet, these authors also used the Balmer lines
and the MgII 4481 line. The Balmer lines, however, are only marginally
sensitive to rotation. In fact, as shown in Figure 8 of
\citet{mucciarelli+14}, a change of rotational velocity from 0 to 200
km s$^{-1}$ produces no significant variations in the line wing
morphology and the full width at half maximum of the Balmer lines. The
Mgb triplet lines are by far more sensitive to rotation than the
Balmer ones, and the estimates they provide are therefore
significantly more precise.  This is also confirmed by the fact that
the difference between the rotational velocities measured by SP14 from
the Mgb triplet alone and the values determined in this work is
reduced: the dispersion of the data with respect to zero decreases
from $\sigma=8.7$ to $\sigma=6.4$ (see Figure
\ref{fig:comparison_SP14}). Indeed, by using the rotational velocities
measured from the Mgb diagnostic alone, the fraction of BSSs spinning
faster than 40 km s$^{-1}$ in the SP14 sample amounts to $\sim$
20-25\% of the total, which is not that different from what we
obtained here ($\sim 28\%$).
 
\begin{figure}[ht!]
\includegraphics[width=9cm, height=9cm]{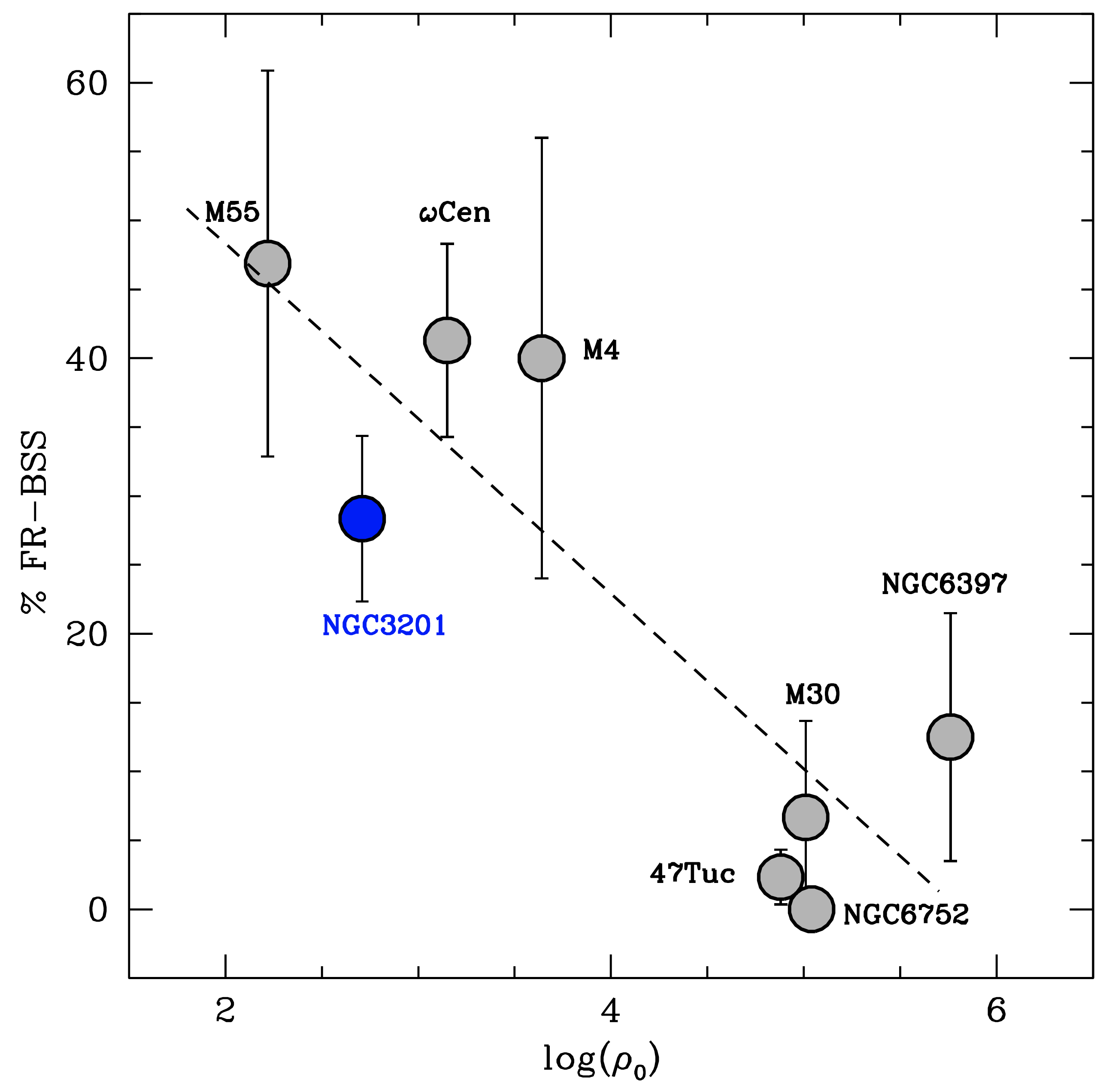}
\centering
\caption{Percentage of FRs (i.e., number of BSSs with $v \sin(i)>40$
  km s$^{-1}$ over the total number of surveyed BSSs) as a function of
  the central density of the parent cluster, for NGC 3201 (blue
  circle) and the other GCs (namely, M55, $\omega$ Centauri, M4, 47
  Tucanae, M30, NGC 6752. and NGC 6397; grey circles) studied in
  \citet{ferraro+23b}.}
\label{fig:total}
\end{figure}

\begin{figure}[ht!]
\includegraphics[width=17cm, height=17cm]{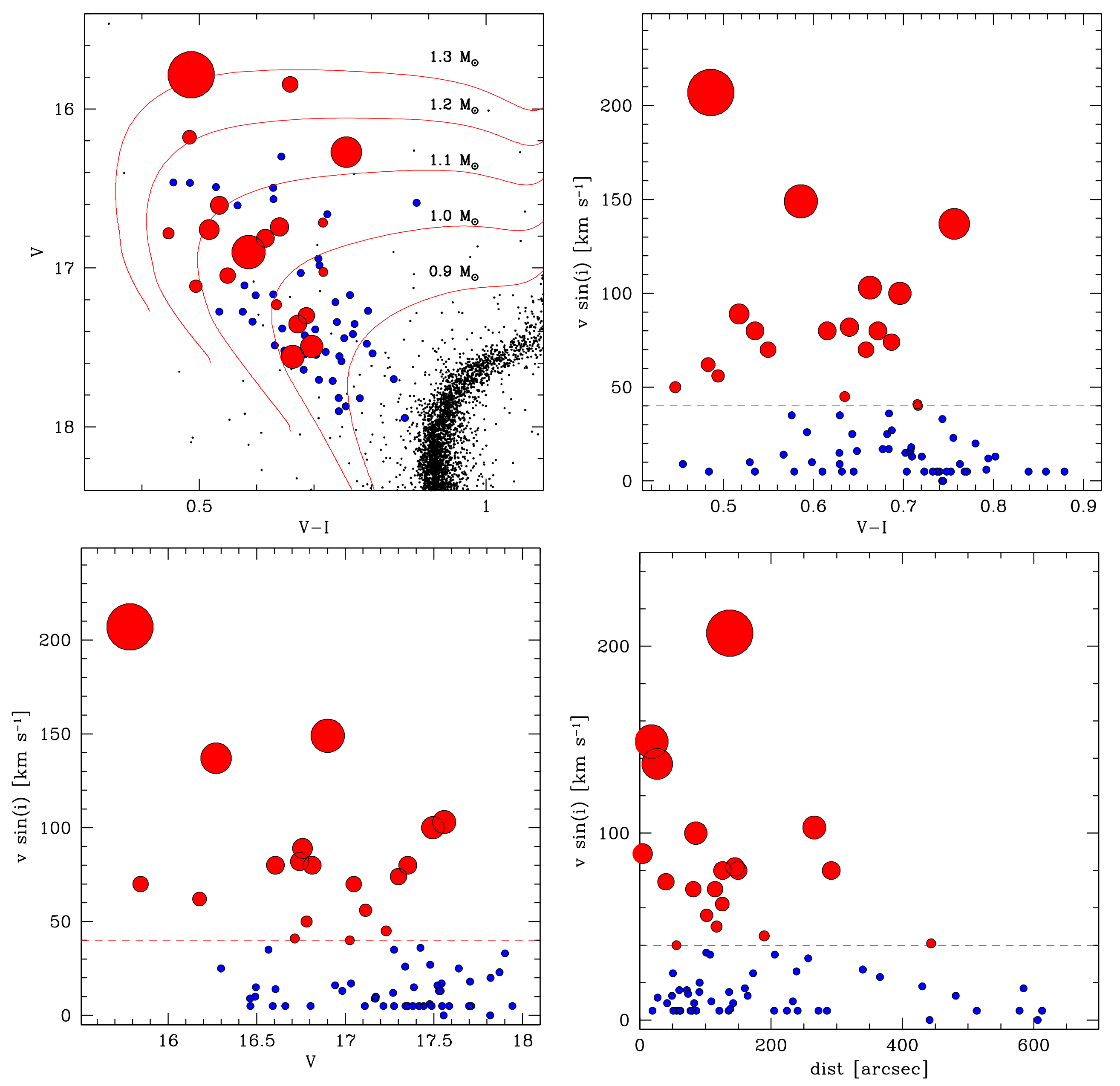}
\centering
\caption{From top-left, to bottom-right: distribution of the surveyed
  BSSs in the CMD of NGC 3201, and rotational velocity distributions
  as a function of the $(V-I)$ color, the $V$ magnitude, and the
  distance from the cluster center. In all the panels, the FRs are
  plotted as red circles with increasing size for increasing
  rotational velocity, while the BSSs spinning at $v \sin(i)<40$ km
  s$^{-1}$ are marked as small blue circles. A set of single-mass
  evolutionary tracks \citep{pietrinferni+21} for stars between 0.9
  and $1.3 M_\odot$ are also over-plotted as red lines in the top-left
  panel.}
\label{fig:4panels}
\end{figure}

\section{Discussion and conclusions}
\label{sec:discussion}
As discussed in the Introduction, the BSSs analyzed in this work were
part of the sample studied in \citet{ferraro+23b}, who found a strong
anti-correlation between the fraction of FRs (defined as BSSs spinning
at more than 40 km s$^{-1}$)\footnote{The 40 km s$^{-1}$ threshold was
adopted from the inspection of the BSS rotational velocity
distributions in 8 GCs. \citet{ferraro+23b} also showed that by
adopting a slightly different threshold (30 or 50 km s$^{-1}$) the
overall results remain unchanged} and the central density of the
parent cluster, suggesting that loose environments are the ideal
habitat for highly rotating BSSs.  This is shown in Figure
\ref{fig:total}, which is a reproduction of Figure 3 of
\citet{ferraro+23b} and where the position of NGC 3201 is highlighted
in blue. As can be seen, the percentage of FRs in NGC 3201 is fully
compatible with that of the other low-density clusters in the sample
(namely, M55, $\omega$ Centauri, and M4), while it is significantly
larger than that measured in the four high-density systems (namely, 47
Tucanae, M30, NGC 6752 and NGC 6397).

High rotational velocities are expected to be a characteristic feature
of recently born BSSs in both the proposed formation scenarios, as the
natural effect of angular momentum accretion during mass transfer
(\citealp{packet+81,demink+13}), and angular momentum conservation
during the contraction phase in the post-collision evolution
\citep[e.g.,][]{benz+87,sarna+96}.  Then, a progressive slow down of
these stars due to the occurrence of some braking mechanisms (like
magnetic braking or disk locking) is predicted, but the efficiencies
of these processes are still unknown \citep{leonard+95, sills+05}. A
few constraints are finally emerging from the analysis of BSSs in open
clusters \citep{leiner+18} and in post core collapse systems
\citep{ferraro+23b}, both indicating a braking timescale of the order
of 1-2 Gyr for both mass-transfer and collisional BSSs.  Following this
scenario, a large rotational velocity can be considered as the
signature of the early phase of BSS evolution, and the percentage of
FRs therefore provides information about the recent BSS formation
activity in the host cluster.  On the other hand, stellar collisions
are expected to be negligible in low-density environments, where the
main BSS formation channel likely is mass transfer in binary
systems. Indeed, the fraction of FRs has been found to also correlate
with the fraction of binaries (see Figure 4 of \citealp{ferraro+23b}).
Hence, the significantly larger fraction of FRs measured in loose than
in dense clusters has been interpreted as the evidence of recent BSS
formation from the evolution of primordial binaries, which are instead
subject to the destructive action of multiple dynamical interactions
in high-density environments.  The case of NGC 3201 should be
interpreted in this context: the measured percentage of FRs ($28\%$)
indicates an intense and recent BSS formation activity from the mass
transfer channel.
 
\begin{figure}[ht!]
\includegraphics[width=16cm, height=8cm]{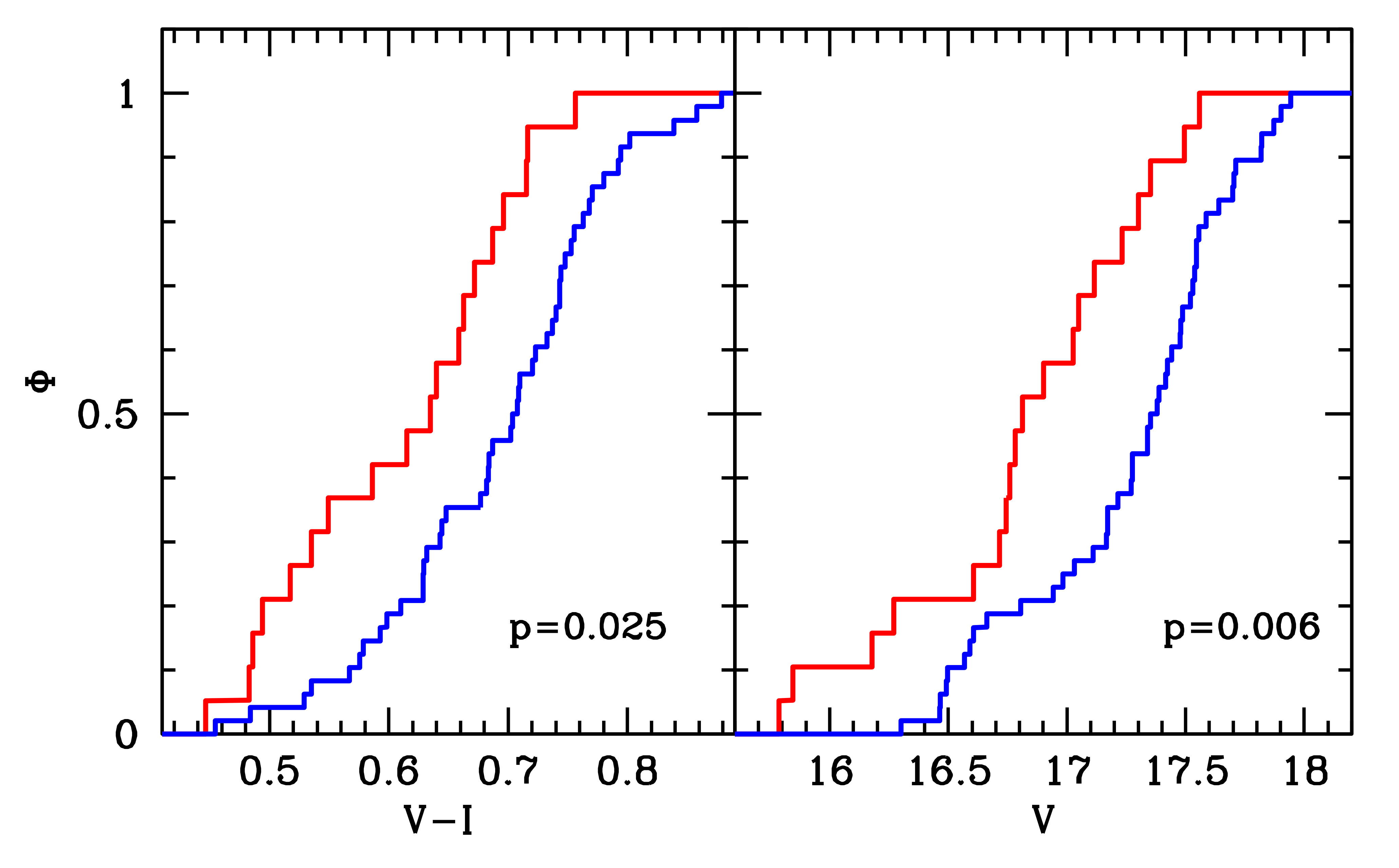}
\centering
\caption{Cumulative distributions of the $(V-I)$ color (left panel)
  and $V$ magnitude (right panel) for the sample of FRs (i.e., BSSs
  with $v \sin(i)\ge 40$ km s$^{-1}$; red lines) and slow rotators
  (blue lines).  The Kolmogorov-Smirnov probability to be extracted
  from the same parent distribution $p$ is marked in each panel.  Note
  that the statistical significance of the Kolmogorov-Smirnov test is
  somehow overestimated in this case due to the need of introducing a
  free parameter (namely the 40 km s$^{-1}$ threshold).}
\label{fig:cumu}
\end{figure}

\begin{figure}[ht!]
\includegraphics[width=9cm, height=9cm]{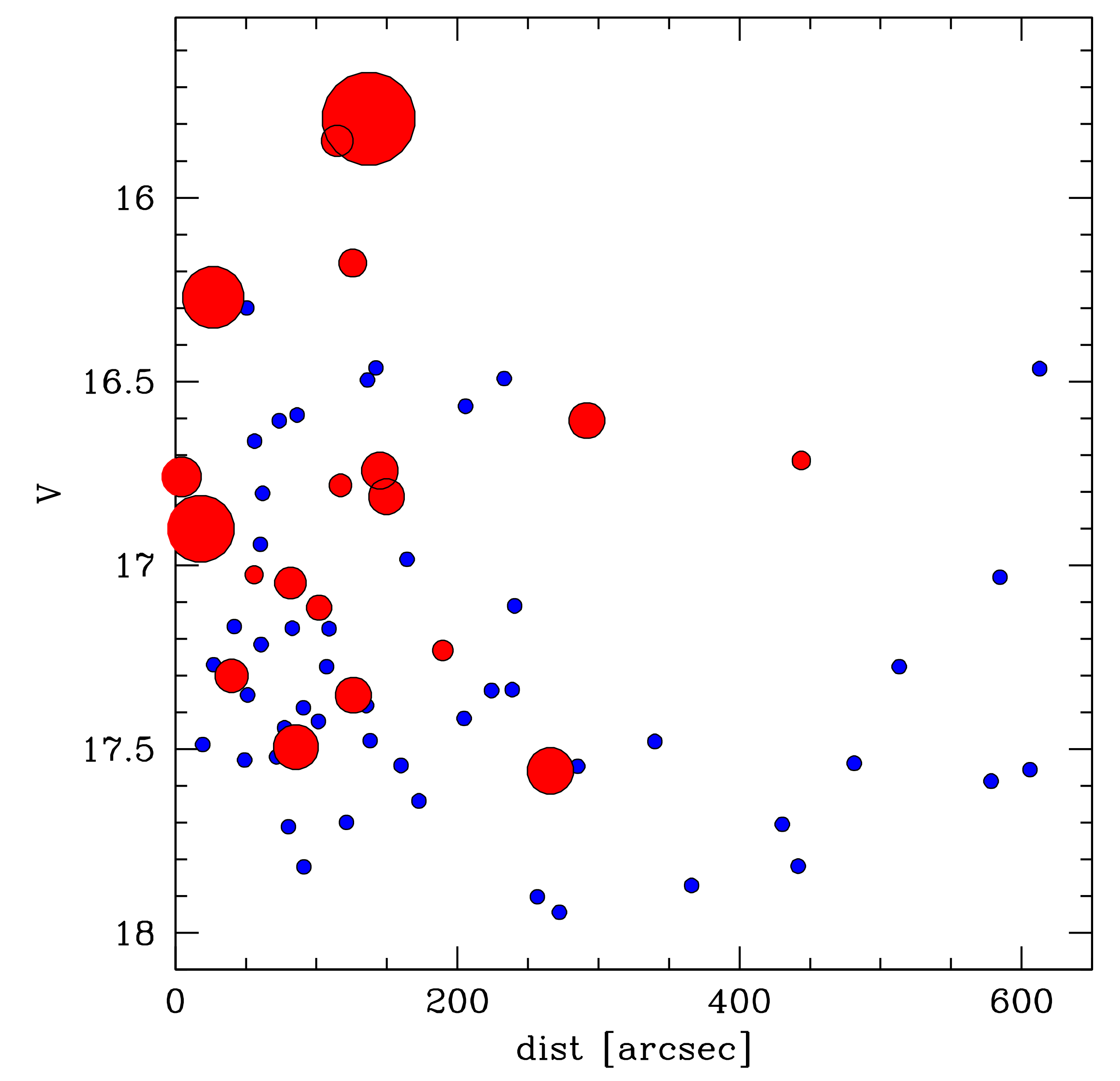}
\centering
\caption{Distribution of magnitude as a function of the distance from
  the cluster centre (in arcseconds) for the observed sample of
  BSSs. Symbols have the same meaning of Fig. \ref{fig:4panels}}.
\label{fig:vdist}
\end{figure}

Additional information can be obtained from the analysis of the CMD
and radial distributions of the observed stars. The top-left panel of
Figure \ref{fig:4panels} shows the CMD of NGC 3201 zoomed in the BSS
region, with the slow rotators ($v \sin(i)<40$ km s$^{-1}$) marked as
small blue circles, and the FRs plotted as red circles of increasing
size for increasing rotational velocity.
%% A set of single-star evolutionary tracks \citep{pietrinferni+21} with
%% different masses (see labels) are also overplotted as red lines.
%% Although it is generally accepted that more luminous BSSs are more
%% massive than their lower luminosity sisters, caution must be used in
%% deriving BSS masses from their luminosity distribution (e.g.,
%% \citealt{geller+11}). However, the analysis performed in
%% \citet{raso+19} found a reasonable agreement between the mass derived
%% from the fitting of the spectral energy distribution, and that
%% obtained from the comparison of the BSS location in the CMD with
%% single-star evolutionary tracks, thus concluding that the latter
%% method generally provides a reasonable first-guess estimate of the BSS
%% mass.
As apparent for the figure, fast and slowly spinning BSSs are not
homogeneously mixed in the CMD: FRs tend to be preferentially located
in the upper-left side of the BSS sequence, while the region closer to
the MS-TO is populated by slow rotators only. This suggests the
existence of a trend between the rotational velocity and the BSS
luminosity and color (temperature) that is worth to be further
investigated.

The dependence of $v \sin(i)$ on the stellar color is shown in the
upper-right panel of Fig. \ref{fig:4panels}.  Although the statistics
is small, not a single FR is observed at $(V-I)>0.75$, and a hint of
decreasing rotational velocity for increasing color (i.e., decreasing
surface temperature) can also be recognized. The tendency for FRs to
preferentially have bluer colors than slowly spinning BSSs is also
confirmed by the cumulative color distributions of the two samples
that is shown in the left panel of Figure \ref{fig:cumu}. The
significance of the detected difference is not very high (of the order
of 2.2$\sigma$). However the same signal seems to emerge (even more
clearly) with the magnitude. In fact the inspection of the bottom-left
panel of Fig. \ref{fig:4panels} shows a tendency for the most luminous
BSSs to be the faster rotators, and no slowly spinning BSSs are
observed at the brightest magnitudes.  Although the statistics is
relatively poor, the comparison between the cumulative luminosity
functions of the fast and slow rotator samples (Fig. \ref{fig:cumu})
seems to confirm this trend.

Thus a general tendency to find fast spinning BSS at higher luminosity
and at bluer colors (higher temperatures) is emerging from this
study. The observed trend might also be read in terms of stellar
  mass (see the set of single-star evolutionary tracks overplotted as
  red lines in the upper-let panel of Fig. \ref{fig:4panels}). Indeed,
  it is generally accepted that more luminous BSSs are more massive
  than their lower luminosity sisters \footnote{Although caution
  must be used in deriving BSS masses from their luminosity
  distribution (e.g., \citealt{geller+11}), a reasonable agreement has
  been found \citep{raso+19} between the mass derived from fitting the
  spectral energy distribution, and that obtained from the comparison
  of the BSS location in the CMD with single-star evolutionary
  tracks.}, and such a connection between mass and luminosity is also
  confirmed by the observed radial distribution (see the bottom-right
  panel of Fig. \ref{fig:4panels} and Figure \ref{fig:vdist}): highly
  spinning BSSs (corresponding to the brighter, more massive, objects)
  are not found in the external regions of the cluster and tend to be
  preferentially located toward the center.  This is consistent with
  an effect of mass segregation. Indeed, according to the value of the
  $A^+_{rh}$ parameter \citep{ferraro+23a}, NGC 3201 has an
  intermediate dynamical age, thus implying that dynamical friction is
  currently effective in making the heavier stars sinking toward the
  cluster center starting from the most massive ones (notably, 13 out
  of 19 FRs, i.e. almost $\sim 70\%$ of the total, appear to be more
  massive than $1 M_\odot$ according to the evolutionary tracks shown
  in the upper-left panel of Fig. \ref{fig:4panels}).

  {\it How can the observed trend between rotational velocity and
    temperature (mass) be interpreted?}  Indeed, a decrease of
  rotational velocity for decreasing surface temperature has been
  observed also in the BSS population of the open cluster NGC 188
  \citep{mathieu+09}.  Under the hypothesis that BSS rapid rotation in
  low-density environments is due to mass transfer, followed by a
  braking timescale of the order 1-2 Gyr, the rapid rotation of more
  massive BSSs in NGC 3201 may indicate that the most recent BSS
  formation activity preferentially involved more massive primordial
  binaries.  Alternatively, it is worth noticing that normal MS stars
  show a similar trend, with an increasing rotation in
  higher-temperature objects ($V-I < 0.6$) that is attributed to their
  reduced convective envelopes, leading to much lengthened spin-down
  times (see, e.g., \citealp{kraft67, vansaders+13}).  Hence, a
  similar process could be active also in the interior of BSSs, and
  this would be consistent with the hypothesis that most of these
  stars in NGC 3201 are generated from the evolution of binary
  systems: in fact, no convective envelope is predicted to develop in
  collisional products \citep{sills+05}.  Although the internal
  structure of BSSs is still poorly constrained, the connection of
  rotation with mass and temperature might be a useful clue for
  improved BSS models.

In conclusion, the statistical significance of the discussed trends is
admittedly low, and many crucial pieces of information are still
missing (including information about the binarity of these systems and
their possible tidal synchronization; for recent results in the field
and open clusters, see, e.g., \citealp{lurie+17, leiner+19}, and
references therein). However, the results presented in this work can
be considered as the starting point for future observational campaigns
and possible clues for improved theoretical models aimed at properly
describing the physical properties of these puzzling stars, and the
role that the host environment can play in setting them.
 
%%%%%%%%%%%%%%%%%%%%%%%%%%%%%%%%%%%%%%%%%%%%%%%%%%%%%%%%%%%%%%%%%%%%%%%%%%%%%%%%%%
%\begin{deluxetable*}{rrrrccccc}[ht!]
%\tablecaption{TABLE CAPTION.}
%\tablewidth{0pt}
%\renewcommand{\arraystretch}{1.2}
%\tablehead{
%\colhead{col1 } & \colhead{ col2 } & \colhead{ col3 } & \colhead{ col4}  &  \colhead{col5}  &
%\colhead{ col6} & \colhead{ col7}
%}
%\startdata
%    &   &  &   &  &  &   \\ 
%    &   &  &   &  &  &   \\ 
 %   &   &  &   &  &  &   \\
%\enddata
%\tablecomments{TABLE COMMENTS.}
%\label{tab_vrot_bin}
%\end{deluxetable*}
%%%%%%%%%%%%%%%%%%%%%%%%%%%%%%%%%%%%%%%%%%%%%%%%%%%%%%%%%%%%%%%%%%%%%%%%%%%%%%%%%%


%%%%%%%%%%%%%%%%%%%%%%%%%%---------APPENDIX
%\appendix
%\section{Dataset list}\label{appebdix}
%%%%%%%%%%%%%%%%%%%%%%%%%%%%%%%%%%%%%%%%%%%%%%%%%%%%%%%%%%%%%%%%%%%

\vskip1truecm
We warmly thank the anonymous referee for the careful revision of the
manuscript and the provided comments, which help improving the
presentation of the results.  This work is part of the project {\it
  Cosmic-Lab (“Globular Clusters as Cosmic Laboratories”)} at the
Physics and Astronomy Department ``A. Righi" of the Bologna University
(http://www.cosmic-lab.eu/ Cosmic-Lab/Home.html). The research was
funded by the MIUR throughout the PRIN-2017 grant awarded to the
project {\it Light-on-Dark} (PI:Ferraro) through contract
PRIN-2017K7REXT.


\newpage
%\bibliography{}{}

\bibliographystyle{aasjournal}



\end{document}